\documentclass[ reprint,
 amsmath,amssymb,
 aps,
]{revtex4-2}

\usepackage{graphicx}
\usepackage{mathptmx}
\usepackage{epstopdf}
\usepackage{dcolumn}
\usepackage{bm}
\usepackage{float}
\usepackage{multirow}
\usepackage[colorlinks=true,
            linkcolor=blue,
            anchorcolor=blue,
            urlcolor=blue,
            citecolor=blue]{hyperref} 

\usepackage{CJK}
\usepackage{flushend}
\usepackage{amsmath}
\usepackage{color}
\usepackage{xcolor}
\usepackage[switch]{lineno}
\usepackage[T1]{fontenc}

\begin{document}
\preprint{APS/123-QED}

\title{The evolution of the chiral symmetry in cesium isotopes}

\author{Duo Chen}
\affiliation{%
	College of Physics, Jilin University, Changchun 130012, China
}%
\author{Jian Li}
\email{E-mail:jianli@jlu.edu.cn}
\affiliation{%
College of Physics, Jilin University, Changchun 130012, China
}%
\author{Rui Guo}
\affiliation{%
	School of Physics, Beihang University, Beijing 100191, China
}%

\begin{abstract}
Following the reports of candidate chiral doublet bands observed in cesium isotopes, the possible chiral candidates and the evolution of three-dimensional rotation in $^{120-134}{\textrm{Cs}}$ are investigated within the microscopic three-dimensional tilted axis cranking covariant density functional theory (3DTAC-CDFT). By investigating the evolution of the polar angle $\theta$ and azimuth angle $\varphi$ as a function of rotational frequency $\hbar\omega$, the transition from the planar rotation to the chiral rotation has been found in $^{121-133}{\textrm{Cs}}$. The corresponding critical rotational frequency $\omega_{\textrm{crit}}$ of the appearance of chiral aplanar rotation decreases as neutron number increases, which can be attributed to the neutrons in $(gd)$ and $(sd)$ shells having smaller angular momentum components along both the short and long axes, and larger components along medium axis, respectively. In comparison, only planar rotation has been obtained in $^{120,134}{\textrm{Cs}}$. With these interpretations, the obtained $I\sim\hbar\omega$ and energy spectra as well as  $B(M1)/B(E2)$ values show reasonable agreement with the available experimental data. In addition, the evolution of quadrupole deformation $\beta$ and triaxial deformation $\gamma$ are also discussed.
\end{abstract}

\maketitle

\section{Introduction}

Chirality has attracted general interest in natural sciences, such as chemistry, biology, and physics. In nuclear physics, chirality, i.e., the spontaneous chiral symmetry breaking, was originally suggested by Frauendorf and Meng in 1997~\cite{FRAUENDORF1997131}. In a triaxially deformed nucleus with the unpaired particle(s) and hole(s) in the high-$j$ orbit, the particle(s) and hole(s) align their angular momentum vectors along the short(s) and long(l) axes, respectively, while the collective core aligns along the medium(m) axis, the total angular momentum vector lies outside the three principal planes in the intrinsic frame (aplanar rotation) with left- or right-handed orientations, which can form a chiral system. The restoration of the spontaneous chiral symmetry breaking in the laboratory frame could give rise to the observation of the so-called chiral doublet bands, i.e., a pair of nearly degenerate $\Delta{I}=1$ bands with the same parity~\cite{FRAUENDORF1997131}. In fact, a pair of $\Delta I = 1$ bands found in $^{134}$Pr with the $\pi{h_{11/2}}\otimes\nu{h_{11/2}}$ configuration observed early in 1996~\cite{PETRACHE1996106} have been reinterpreted in Ref.~\cite{FRAUENDORF1997131} as a candidate for chiral doubling. Thereafter, in 2001 similar low-lying doublet bands were reported in $_{55}$Cs, $_{57}$La, and $_{61}$Pm $N = 75$ isotones of $^{134}$Pr, and an island of chiral rotation was suggested in the $A\sim130$ mass regionin~\cite{PhysRevLett.86.971}. It should be noted that clearly identifing the chiral motion is difficult, for example the chirality in $^{134}$Pr aforementioned was later rejected in 2006~\cite{PhysRevLett.96.112502, PhysRevLett.96.052501}, and in 2011~\cite{PhysRevC.84.044302} other chiral candidates in $^{134}$Pr were proposed.


Since this pioneering work, a lot of experimental and theoretical works related to nuclear chirality have been done, see Refs.~\cite{Meng_2010, Meng2014, Meng_2016, Raduta2016}, for a brief review. Thus nuclear chirality becomes a hot topic and related issues such as M$\chi$D (multiple chiral doublet bands in one nucleus)~\cite{PhysRevC.73.037303, PhysRevLett.110.172504, PhysRevLett.113.032501, PhysRevC.97.041304(R), PhysRevC.83.037301}, the nuclear Chirality-Parity (ChP) violation~\cite{PhysRevLett.116.112501, Wang2020} and so on have been widely discussed.

Up to now, more than $60$ candidate chiral nuclei have been reported in the ${A}\ {\sim}\ {80}, {100}, {130}, {160}, {\textrm{and }}{190}$ mass regions of the nuclear chart, see, e.g., data tables~\cite{XIONG2019193}. In particular, the reported candidate chiral nuclei in ${A}{\sim}{130}$ mass region form a large chiral island, where the cesium isotopes have the most chiral candidates. So far, chiral doublet bands have been observed in the odd-odd nuclei $^{122,124,126,128,130,132}$Cs~\cite{PhysRevLett.86.971, PhysRevLett.97.172501, U_2005, GRODNER201146, Dong_2009, PhysRevC.96.051303(R), PhysRevC.68.024318} with configuration ${{\pi{{h}^{1}_{11/2}}}{\otimes}{\nu{{h}^{-1}_{11/2}}}}$, and corresponding chirality are also well explained by different theoretical models~\cite{U_2005, PhysRevC.82.027303, PhysRevC.75.024309, Shirinda2012, BHAT2014150, PhysRevLett.97.172501, PhysRevC.103.024327, QiBin_2009, PhysRevC.96.051303(R), PhysRevC.82.034328, PhysRevC.67.044319, PhysRevLett.93.172502, CHEN2018211, PhysRevC.83.054308, PhysRevC.68.044324, PhysRevC.73.054308, PhysRevC.68.024318, PhysRevC.72.024315}. In fact, chiral doublet bands were first identified in the ${A}\ {\sim}\ {130}$ nuclei including $^{130}{\textrm{Cs}}$~\cite{PhysRevLett.86.971}. The chiral doublet bands found in $^{128}{\textrm{Cs}}$ was proposed as the best example to reveal the chiral symmetry breaking~\cite{PhysRevLett.97.172501}, and the succeeding $g$-factor measurements can give important information on the relative orientation of the three angular momentum vectors~\cite{PhysRevLett.120.022502, PhysRevC.106.014318}. Meanwhile, the lifetimes of excited states belonging to the chiral partner bands have been reported in $^{126}{\textrm{Cs}}$, which is the first time a large set of experimental transition probabilities are in qualitative agreement with all selection rules predicted for the strong chiral symmetry breaking limit~\cite{GRODNER201146}. Therefore, it is naturally interesting to find nuclear chirality in other cesium isotopes. In neighbouring odd-odd isotopes $^{120,134}{\textrm{Cs}}$, only one rotational band with the configuration ${{\pi{{h}^{1}_{11/2}}}{\otimes}{\nu{{h}^{-1}_{11/2}}}}$ was reported~\cite{PhysRevC.58.1849, PhysRevC.67.044319}. In the heavier $^{134}{\textrm{Cs}}$, the experimental observations~\cite{PhysRevC.67.044319, PhysRevC.84.041301(R)} indicate that the structure of the band is distinctly different from its lighter odd-odd isotopes, and was interpreted as a possible magnetic rotation band~\cite{PhysRevC.84.041301(R)}. Therefore, it is worthwhile to recheck the rotational modes in $^{120,134}{\textrm{Cs}}$ and the possible boundaries of the cesium isotopes.

For the odd-$A$ cesium isotopes, doublet bands involving the configuration component ${{\pi{{h}^{1}_{11/2}}}{\otimes}{\nu{{h}^{-1}_{11/2}}}}$ with the third neutron quasi-particle or hole by the ${g_{7/2}/d_{5/2}}$ or ${s_{1/2}/d_{3/2}}$ orbits in $^{121,123,125,127,129,131}{\textrm{Cs}}$~\cite{{Singh2006}, {PhysRevC.78.034313}, {PhysRevC.79.044317}, {PhysRevC.42.890}, {etde_20450471}, {Singh2005}} have been observed. Theoretically, the constrained triaxial relativistic mean-field (RMF) approach~\cite{PhysRevC.97.034306, PhysRevC.100.034328} has been performed to study the possible chirality in odd-$A$ cesium isotopes $^{125,127,129,131}{\textrm{Cs}}$, and the existence of chirality is demonstrated and expected based on different high-$j$ particle-hole configurations and triaxial deformations. In addition, a rotational band with the configuration component ${{\pi{{h}^{1}_{11/2}}}{\otimes}{\nu{{h}^{-1}_{11/2}}}{{(sd)}^{-1}}}$ has been observed in $^{133}{\textrm{Cs}}$~\cite{PhysRevC.95.064320}. Therefore, it is also interesting to investigate the chirality in other odd-$A$ cesium isotopes, i.e., $^{121,123,133}{\textrm{Cs}}$.


In order to describe the chiral geometry microscopically, we adopt the three-dimensional tilted axis cranking based on the covariant density functional theory (3DTAC-CDFT). This approach has been developed~\cite{ZHAO20171} and used to investigate the nuclear chirality in $^{106}{\textrm{Rh}}$~\cite{ZHAO20171}, $^{106}{\textrm{Ag}}$~ \cite{PhysRevC.99.054319}, $^{135}{\textrm{Nd}}$~\cite{PENG2020135795} and $^{102-107}{\textrm{Rh}}$~\cite{PhysRevC.105.044318}. In these studies, the existence of the critical frequency $\omega_{cirt}$ corresponding to the transition from planar to aplanar rotation is supported. For example, in Ref.~\cite{PENG2020135795}, a classical Routhian was extracted by modeling the motion of the nucleons in rotating mean field as the interplay between the single-particle motions of several valence particle(s) and hole(s) and the collective motion of a core-like part. This classical Routhian gives qualitative agreement with the 3DTAC-CDFT result of $\omega_{cirt}$. In this paper, following the similar procedures outlined in Refs.~\cite{{ZHAO20171}, {PhysRevC.99.054319}, {PENG2020135795}, {PhysRevC.105.044318}}, we will pay attention to chirality in $^{120-134}{\textrm{Cs}}$ within 3DTAC-CDFT.

\begin{figure*}[ht]
\centering
\includegraphics[width=14cm]{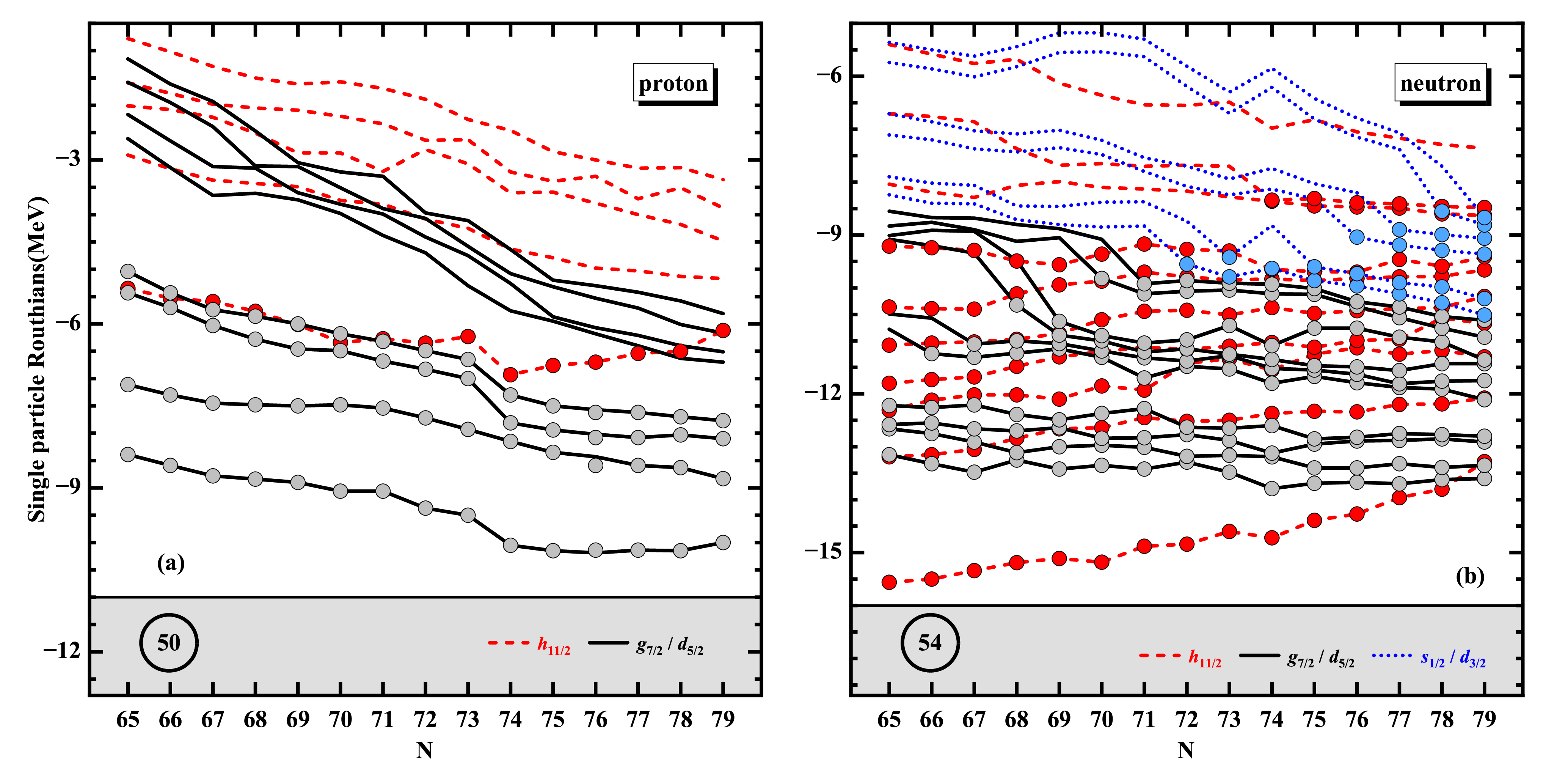}
\caption{\label{fig:single_Routhian} (color online) Single proton (a) and neutron (b) Routhians near the Fermi surface in $^{120-134}{\textrm{Cs}}$ as a function of the neutron number based on the fixed configuration when the rotational frequency is $0.3~{\textrm{MeV}}$. The single proton and neutron levels belonging to the orbits ${g_{7/2}}/{d_{5/2}}$ and ${s_{1/2}}/{d_{3/2}}$ are marked by solid black and blue lines, respectively, and the levels belonging to $h_{11/2}$ orbit with negative parity are indicated by dashed red lines. The red, grey and blue filled circles indicate the occupied levels in the orbits $h_{11/2}$, ${g_{7/2}}/{d_{5/2}}$ and ${s_{1/2}}/{d_{3/2}}$.}\label{fig1}
\end{figure*}

\section{Theoretical framework}

The 3DTAC-CDFT based on the zero-range point-coupling interaction has been successfully applied to describe the chiral rotation~\cite{{ZHAO20171}, {PhysRevC.99.054319}, {PENG2020135795}, {PhysRevC.105.044318}}. The detailed formalism of the 3DTAC-CDFT can be found in Ref.~\cite{ZHAO20171}. Here, a brief introduction is presented.

The $\textrm{CDFT}$ starts from a Lagrangian, and the corresponding Kohn-Sham equations have the form of a Dirac equation with effective fields $S$ and $V$ derived from this Lagrangian~\cite{{Reinhard1989}, {RING1996193}, {VRETENAR2005101}, {MENG2006470}, {doi:10.1142/9872}}. In the $\textrm{3DTAC}$ method, these fields are calculated in the intrinsic frame rotating with a constant angular velocity vector $\boldsymbol{\omega}$, pointing in an arbitrary direction in space~\cite{ZHAO20171}

\begin{equation}
[\boldsymbol{\alpha} \cdot(\boldsymbol{p}-\boldsymbol{V})+\beta(m+S)+V-\boldsymbol{\omega} \cdot \hat{\boldsymbol{J}}] \psi_{k}=\epsilon_{k} \psi_{k}.
\label{eq1}
\end{equation}

Here, ${\widehat{\boldsymbol{J}}}$ is the total angular momentum of the nucleon spinors, and the relativistic fields $S$, $V$, and $\bm{V}$ read

\begin{equation}
\begin{aligned}
& S(\boldsymbol{r}) =\alpha_S \rho_S+\beta_S \rho_S^2+\gamma_S \rho_S^3+\delta_S \triangle \rho_S, \\
& V^0(\boldsymbol{r}) =\alpha_V \rho_V+\gamma_V \rho_V^3+\delta_V \triangle \rho_V+\tau_3 \alpha_{T V} \rho_{T V} \\
& \ \ \ \ \ \ \ \ \ \ +\tau_3 \delta_{T V} \triangle \rho_{T V}+e A^0, \\
& V(\boldsymbol{r}) =\alpha_V \boldsymbol{j}_V+\gamma_V\left(\boldsymbol{j}_V\right)^3+\delta_V \triangle \boldsymbol{j}_V+\tau_3 \alpha_{T V} \boldsymbol{j}_{T V} \\
& \ \ \ \ \ \ \ \ \ \ +\tau_3 \delta_{T V} \triangle \boldsymbol{j}_{T V}+e \boldsymbol{A}.
\label{eq2}
\end{aligned}
\end{equation}

The $S$, $V$, and $\bm{V}$ are the relativistic scalar field, the time-like component of vector field, and the space-like components of vector field, respectively, which are in turn coupled with the nucleon densities and current distributions. In the above equation, the $\rho_S$, $\rho_V$, $\rho_{TV}$, $j_V$, $j_{TV}$ represent various densities and currents, and the $\alpha_S$, $\beta_S$, $\gamma_S$, $\delta_S$, $\alpha_V$, $\gamma_V$, $\delta_V$, $\alpha_{TV}$, $\delta_{TV}$ are nine parameters of the Lagrangian. The Dirac equation is solved in a set of three-dimensional harmonic oscillator basis iteratively, and one finally obtains the single-nucleon spinors ${\psi_{k}}$, the single-particle Routhians ${\epsilon_{k}}$, the total energies, the expectation values of the angular momenta, transition probabilities, and so on. Using the semiclassical cranking condition $\langle\widehat{\boldsymbol{J}}\rangle\cdot\langle\widehat{\boldsymbol{J}}\rangle=I(I+1)$, one can relate the magnitude of the angular velocity $\boldsymbol{\omega}$ to the angular momentum quantum number $I$. Meanwhile, the orientation of $\boldsymbol{\omega}$ is determined self-consistently by minimizing the total Routhian.

From the Dirac equation, the physical observables, including the quadrupole moments and magnetic moments, and the electromagnetic transition probabilities $B(M1)$ and $B(E2)$, can be calculated.

The quadrupole moments $Q_{20}$ and $Q_{22}$ are

\begin{equation}
\begin{aligned}
Q_{20} & =\sqrt{\frac{5}{16 \pi}}\left\langle 3 z^2-r^2\right\rangle = \frac{3 A}{4 \pi} R_0^2 a_{20}, \\
Q_{22} & =\sqrt{\frac{15}{32 \pi}}\left\langle x^2-y^2\right\rangle= \frac{3 A}{4 \pi} R_0^2 a_{22}.
\label{eq3}
\end{aligned}
\end{equation}

The atomic quadrupole deformation parameters $\beta$ and triaxial deformation parameters $\gamma$ can be obtained

\begin{equation}
\begin{aligned}
\beta = \sqrt{a_{20}^2+2 a_{22}^2},\ \ \ \ \ \gamma = \arctan \left[\sqrt{2} \frac{a_{22}}{a_{20}}\right].\label{eq4}
\end{aligned}
\end{equation}

The nuclear magnetic moment in units of the nuclear magneton is given by,

\begin{equation}
\begin{aligned}
\boldsymbol{\mu} & =\sum_{k>0} n_k \int d^3 r\left[\frac{m c^2}{\hbar c} q \psi_k^{\dagger} \boldsymbol{r} \times \boldsymbol{\alpha} \psi_k+\kappa \psi_k^{\dagger} \beta \boldsymbol{\Sigma} \psi_k\right],
\label{eq5}
\end{aligned}
\end{equation}

where the charge $q$ is 1 for protons and 0 for neutrons in units of $e$, $\kappa$ is the nucleon anomalous gyromagnetic factor, ${{\kappa}_{p} = {1.793}}$, ${{\kappa}_{n} = {-1.193}}$, $\emph{\textbf{L}}$ and $\boldsymbol{\Sigma}$ are respectively the orbital angular momentum and spin~\cite{PhysRevC.82.054319}. In the following calculations, the ratio $m c^2/\hbar c$ in Eq.~\ref{eq5} is taken as 1, as Refs.~\cite{PhysRevC.96.054324,ZHAO20171}

The transition probabilities $B(M1)$ and $B(E2)$ are calculated in the semiclassical approximation,

\begin{equation}
\begin{aligned}
B(M 1)= & \frac{3}{8 \pi}\left\{\left[-\mu_z \sin \theta+\cos \theta\left(\mu_x \cos \varphi+\mu_y \sin \varphi\right)\right]^2\right. \\
& \left.+\left(\mu_y \cos \varphi-\mu_x \sin \varphi\right)^2\right\}, \\
B(E 2)= & \frac{3}{8}\left[Q_{20}^p \sin ^2 \theta+\sqrt{\frac{2}{3}} Q_{22}^p\left(1+\cos ^2 \theta\right) \cos 2 \varphi\right]^2 \\
& +\left(Q_{22}^p \cos \theta \sin 2 \varphi\right)^2,
\label{eq6}
\end{aligned}
\end{equation}

where $Q^{p}_{20}$ and $Q^{p}_{22}$ are the quadrupole moments of protons, and $\theta$ and $\varphi$ are the orientation angles of the total angular momentum ${\widehat{\boldsymbol{J}}}$ in the intrinsic frame.

In the present work, we adopt the point-coupling Lagrangian PC-PK1~\cite{PhysRevC.82.054319} and the calculations are free of additional parameters. The Dirac equation Eq.~\ref{eq1} is solved in a three-dimensional Cartesian harmonic oscillator basis with 10 major shells. The pairing correlation is neglected in the calculations, but one has to bear in mind that the pairing correlation might have some influences on the descriptions of critical frequency~\cite{PhysRevC.73.054308} as well as the total angular momentum and $B(M1)$ values~\cite{PhysRevC.92.034319, PhysRevC.87.024314}.

\section{Results and discussion}


\begin{figure*}[ht]
\centering
\includegraphics[width=14cm]{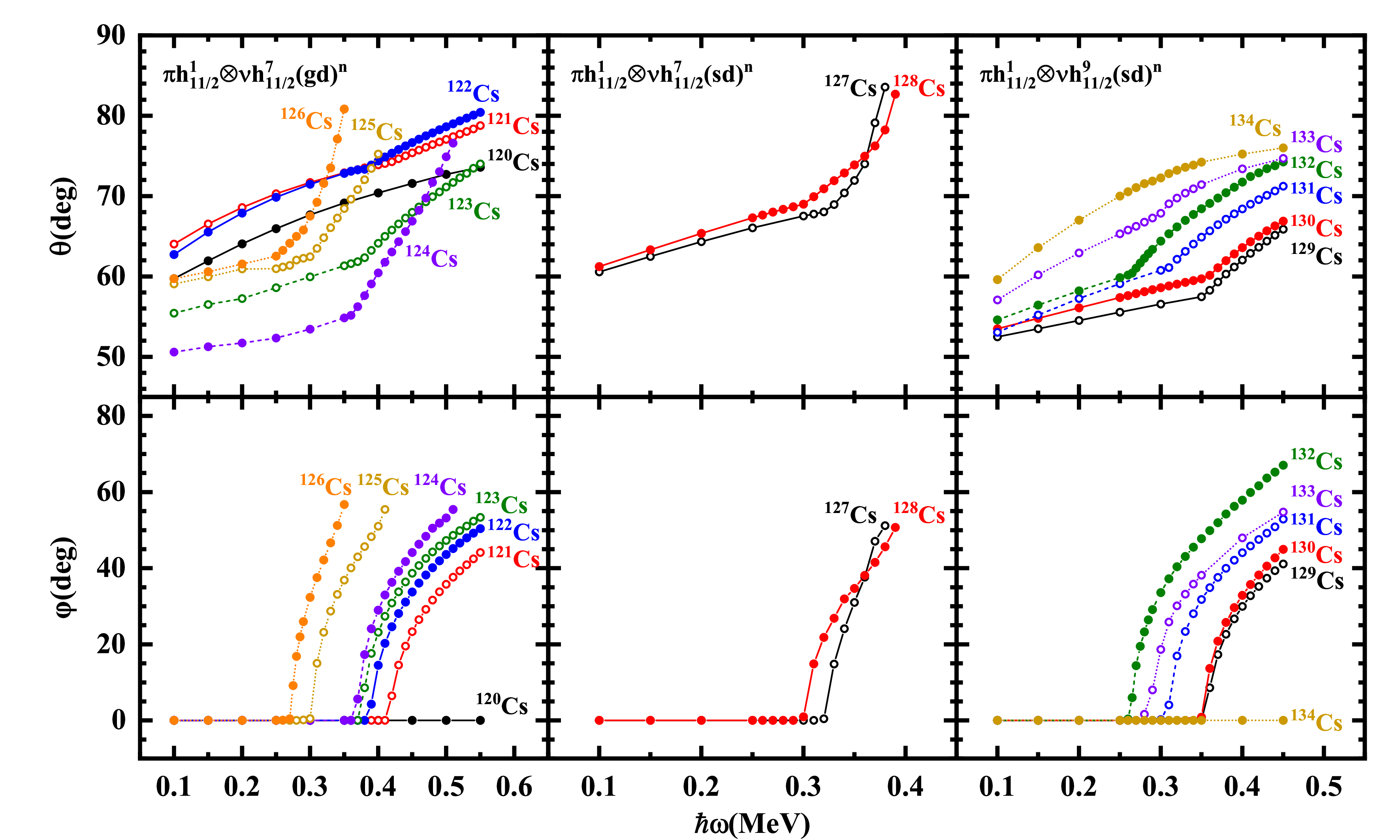}
\caption{\label{fig:thera,phi} (color online) The evolution of the polar angle $\theta$ and azimuth angle $\varphi$ for the total angular momentum $J$ as driven by the increasing rotational frequency $\hbar\omega$ for configurations ${{\pi{{h}^{1}_{11/2}}}{{g}^{4}_{7/2}}{\otimes}{\nu{{h}^{7}_{11/2}}}{{(gd)}^n}{(n=8-14)}}$ in ${^{120-126}{\textrm{Cs}}}$, ${{\pi{{h}^{1}_{11/2}}}{{g}^{4}_{7/2}}{\otimes}{\nu{{h}^{7}_{11/2}}}{{(sd)}^n}{(n=1-2)}}$ in ${^{127-128}{\textrm{Cs}}}$ and ${{\pi{{h}^{1}_{11/2}}}{{g}^{4}_{7/2}}{\otimes}{\nu{{h}^{9}_{11/2}}}{{(sd)}^n}{(n=1-6)}}$ in ${^{129-134}{\textrm{Cs}}}$.}\label{fig2}
\end{figure*}

In the present study for $^{120-134}{\textrm{Cs}}$, two
quasi-particle configurations ${{\pi{{h}^{1}_{11/2}}}{\otimes}{\nu{{h}^{-1}_{11/2}}}}$
and three quasi-particle configurations ${{\pi{{h}^{1}_{11/2}}}{\otimes}{\nu{{h}^{-1}_{11/2}}}{{(gd)}^{-1}}\ {\textbf{\&}}\ {\pi{{h}^{1}_{11/2}}}{\otimes}{\nu{{h}^{-1}_{11/2}}}{{(sd)}^{1}}}$ are assigned for odd-odd nuclei and the odd-$A$ nuclei, respectively, similar to the earlier studies in Refs.~\cite{{PhysRevLett.86.971}, {PhysRevLett.97.172501}, {U_2005}, {GRODNER201146}, {Dong_2009}, {PhysRevC.96.051303(R)}, {PhysRevC.68.024318}, {Singh2006}, {PhysRevC.78.034313}, {PhysRevC.79.044317}, {PhysRevC.42.890}, {PhysRevC.97.034306}, {PhysRevC.100.034328}, {PhysRevC.58.1849}, {etde_20450471}, {MOON2000343}, {PhysRevC.95.064320}, {PhysRevC.67.044319}}. In the calculations, an unpaired proton is always occupying the bottom of the $h_{11/2}$ orbit and the occupation of the valence neutrons in $h_{11/2}$ orbit are fixed by tracing the single particle levels with the increasing frequencies as shown in Fig.~\ref{fig1}, while other nucleons are treated self-consistently by filling the orbits according to their energies from the bottom of the well. The $(gd)$ represents the low-$j$ neutron orbits $1g_{7/2}$ and $2d_{5/2}$, and $(sd)$ represents $2d_{3/2}$ and $3s_{1/2}$, as it is difficult to distinguish them as a strong mixing with each other. Note that the six or eight neutrons in $h_{11/2}$ orbit are antialigned, and the corresponding neutron configuration is ${\nu{{h}^{-1}_{11/2}}}$. In this way, the valence nucleon configurations of ${^{120-126}{\textrm{Cs}}}$ within the ${\textrm{3DTAC-CDFT}}$ calculations are ${{\pi{{h}^{1}_{11/2}}}{{g}^{4}_{7/2}}{\otimes}{\nu{{h}^{7}_{11/2}}}{{(gd)}^n}{(n=8-14)}}$, of ${^{127-128}{\textrm{Cs}}}$ are ${{\pi{{h}^{1}_{11/2}}}{{g}^{4}_{7/2}}{\otimes}{\nu{{h}^{7}_{11/2}}}{{(sd)}^n}{(n=1-2)}}$, and of ${^{129-134}{\textrm{Cs}}}$ are ${{\pi{{h}^{1}_{11/2}}}{{g}^{4}_{7/2}}{\otimes}{\nu{{h}^{9}_{11/2}}}{{(sd)}^n}{(n=1-6)}}$.

In the 3DTAC-CDFT calculations, the orientation of the angular velocity $\boldsymbol{\omega}$ with respect to the principal axis can be determined in a self-consistent manner either by minimizing the total Routhian or by requiring that $\boldsymbol{\omega}$ is parallel to the total angular momentum $J$ at a fixed $\boldsymbol{\omega}$ value. Here, the polar angle $\theta$ and the azimuth angle $\varphi$ are used to denote the direction of the angular velocity ${\boldsymbol{\omega}}={\omega}{({\sin{\theta}}{\cos{\varphi}},{\sin{\theta}}{\sin{\varphi}},{\cos{\varphi}})}$. The $\theta$ is the angle between the angular velocity $\boldsymbol{\omega}$ and the long $(l)$ axis, while $\varphi$ is the angle between the projection of $\boldsymbol{\omega}$ onto the short-medium $(sm)$ plane and the short $(s)$ axis. It has been noted that $\varphi$ can be used to characterize the chirality of a rotational system~\cite{PhysRevC.87.024314}.

To examine the possible presence of chiral geometry, the self-consistently obtained orientation angles $\theta$ and $\varphi$ of the total angular momentum $J$ in the intrinsic frame are shown as a function of the rotational frequency ${\hbar}{\omega}$ in Fig.~\ref{fig2}. The polar angle $\theta$ in ${^{120-134}{\textrm{Cs}}}$ has similar behavior, i.e., they increase with the rotational frequency. Nevertheless, the polar angle $\theta$ is always larger than ${50^\circ}$ in all cesium isotopes. It is attributed to that the angular momentum alignment along the $s$ axis coming from the proton particles in the ${h}_{11/2}$ orbit is much larger than that along the $l$ axis from the neutron holes in the ${h}_{11/2}$ orbit~(cf. Fig.~\ref{fig4}). For comparison, the azimuth angle $\varphi$ for $^{121-133}{\textrm{Cs}}$ is zero at low rotational frequencies, corresponding to the planar rotation in the $sl$ plane. Above the limited rotational frequency corresponding to the so-called critical frequency ${\omega}_\textrm{crit}$ of chiral rotation \cite{{PhysRevC.73.054308}, {PhysRevLett.93.052501}}, the values of $\varphi$ become nonzero and it results in the transition from planar to aplanar rotation. Note that the kinks appear at the curves of $\theta$ for $^{121-133}{\textrm{Cs}}$ with the critical frequency.

The azimuth angle $\varphi$ is always zero for ${^{120,134}{\textrm{Cs}}}$ corresponding to the planar rotation in the $sl$ plane, even the rotational frequency goes up to ${\hbar}{\omega}=0.55$ MeV for ${^{120}{\textrm{Cs}}}$ and ${\hbar}{\omega}=0.45$ MeV for ${^{134}{\textrm{Cs}}}$. By increasing rotational frequency, the results can not be converged for the configuration of ${{\pi{{h}^{1}_{11/2}}}{{g}^{4}_{7/2}}{\otimes}{\nu{{h}^{7}_{11/2}}}{{(gd)}^8}}$ in ${^{120}{\textrm{Cs}}}$ and ${{\pi{{h}^{1}_{11/2}}}{{g}^{4}_{7/2}}{\otimes}{\nu{{h}^{9}_{11/2}}}{{(sd)}^6}}$ in ${^{134}{\textrm{Cs}}}$. Namely, it might mean that ${\hbar}{\omega}_\textrm{crit}$ in the present 3DTAC-CDFT calculations is larger than ${\hbar}{\omega}=0.55$ MeV for ${^{120}{\textrm{Cs}}}$ and ${\hbar}{\omega}=0.45$ MeV for ${^{134}{\textrm{Cs}}}$ even if it exists, which corresponds to ${I}\ {\approx}\ {24}\ {\hbar}$, out of the spin range observed currently~\cite{etde_20450471, PhysRevC.67.044319, PhysRevC.84.041301(R)}. Thus, the present theoretical analysis does not support chirality exsiting in $^{120,134}{\textrm{Cs}}$.

\begin{figure}[h]
\centering
\includegraphics[height=8cm]{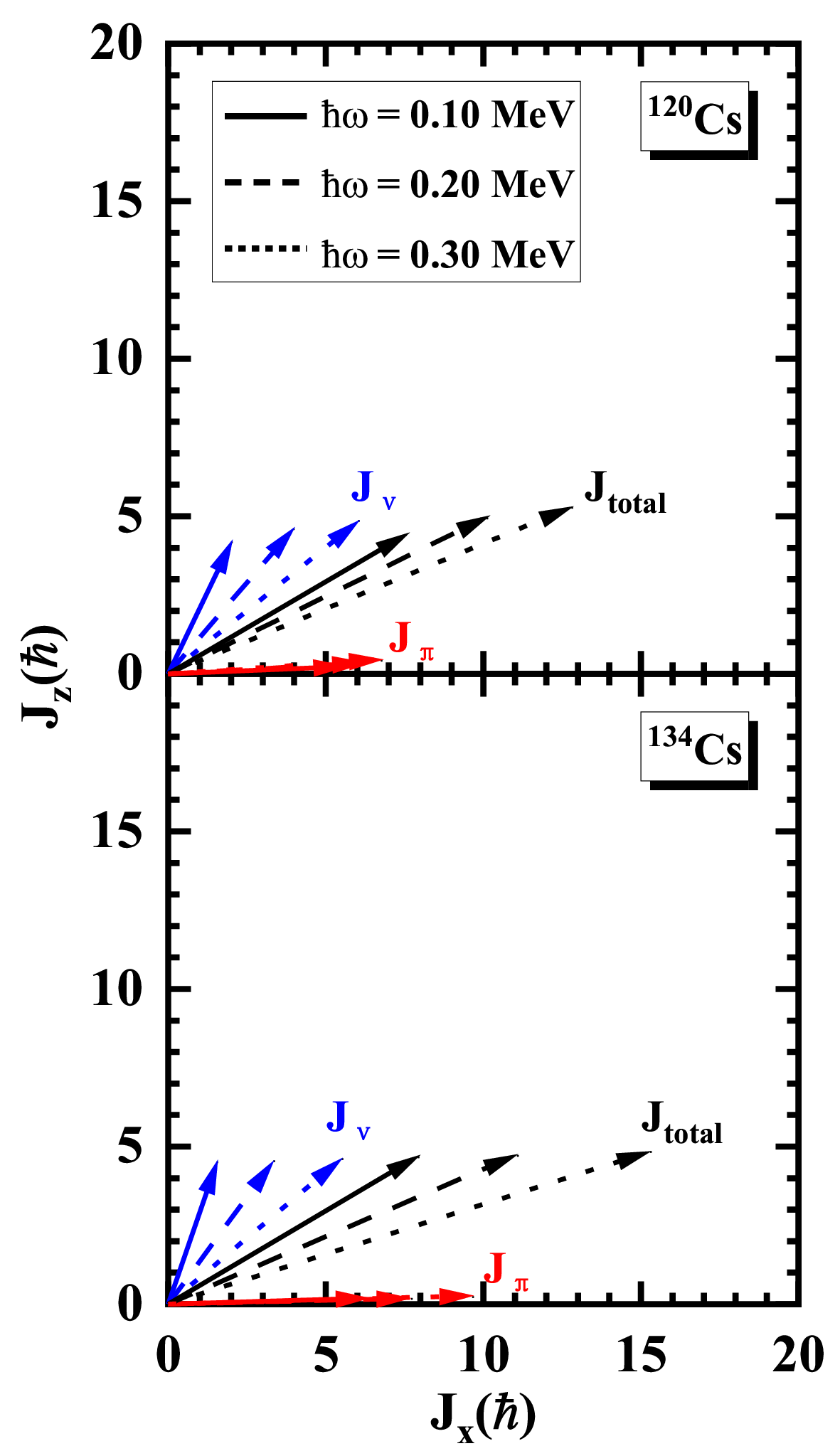}
\caption{\label{fig:angle} (color online) Composition of the total angular momentum at different rotational frequencies in 3DTAC-CDFT calculations for ${^{120}{\textrm{Cs}}}$ and ${^{134}{\textrm{Cs}}}$ with the configuration ${{\pi{{h}^{1}_{11/2}}}{\otimes}{\nu{{h}^{-1}_{11/2}}}}$.}\label{fig3}
\end{figure}

In addition, the previous study indicates a possible MR nature for this configuration ${{\pi{{h}^{1}_{11/2}}}{\otimes}{\nu{{h}^{-1}_{11/2}}}}$ in $^{134}$Cs based on the decreasing trend of $B(M1)$~\cite{PhysRevC.84.041301(R)}. Composition of the total angular momentum at different rotational frequencies in 3DTAC-CDFT calculations for ${^{120}{\textrm{Cs}}}$ and ${^{134}{\textrm{Cs}}}$ with the configuration ${{\pi{{h}^{1}_{11/2}}}{\otimes}{\nu{{h}^{-1}_{11/2}}}}$ have been given in Fig.~\ref{fig3}. In the present 3DTAC-CDFT calculations for the $^{120,134}$Cs, as the four proton particles in the ${g}_{7/2}$ orbital are paired, the proton angular momentum comes mainly from the particles in the ${h}_{11/2}$ orbit, which aligns along the $s$ axis. The neutron hole in the ${h}_{11/2}$ orbit and neutrons in low-$j$ orbits $(gd)/(sd)$ give substantial contributions to the neutron angular momentum, which leads to large components both in $l$ axis and $s$ axis. Higher spin states in the band are created by aligning the neutron angular momentum toward the $s$ axis, while the proton angular momentum keeps unchanged in the $s$ axis. Thus, it is probably the planar rotation in $^{120,134}$Cs.

\begin{figure}[h]
\centering
\includegraphics[width=8cm]{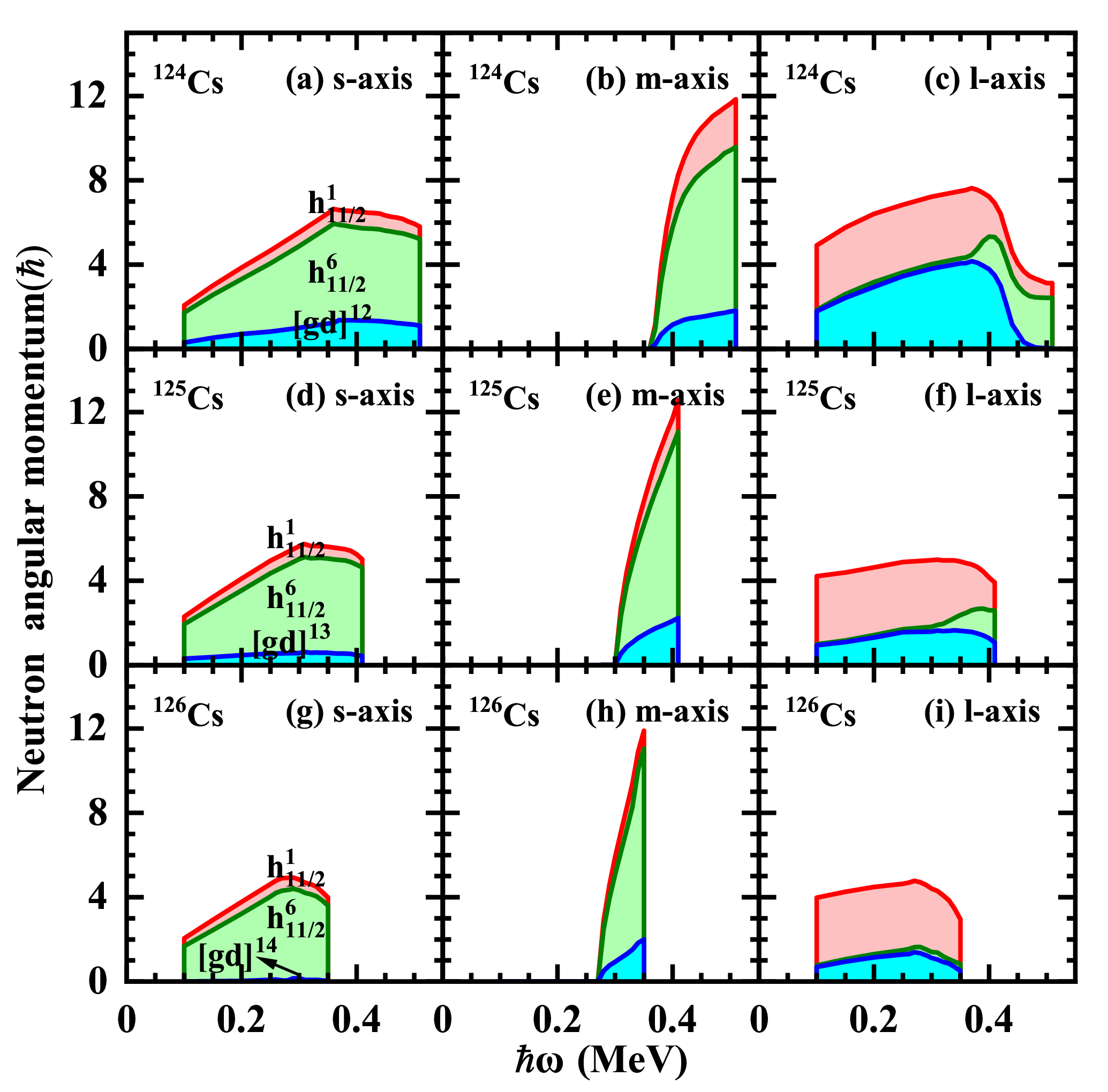}
\caption{\label{fig:angle} (color online) Contributions of the valence neutrons in the ${h}_{11/2}$, $(gd)$ and $(sd)$ orbits to the total angular momenta along the $s$, $m$ and $l$ axis in ${^{124,125,126}{\textrm{Cs}}}$.}\label{fig4}
\end{figure}

The chirality in nuclei with stable triaxial deformation is due to the aplanar rotation formed by the valence particle(s), valence hole(s), and collective core angular momentum vectors~\cite{FRAUENDORF1997131}. Therefore, an important work is the transition point from planar to chiral rotation, i.e., the critical frequency ${\omega}_\textrm{crit}$. As shown in Fig.~\ref{fig2}, the calculated critical frequencies ${\omega}_\textrm{crit}$ all decrease with the increasing mass number for three different configurations in ${^{120-126}{\textrm{Cs}}}$, ${^{127,128}{\textrm{Cs}}}$ and ${^{129-134}{\textrm{Cs}}}$. To understand this interesting behavior, one has to analyze the angular momentum geometry. In 3DTAC-CDFT calculations, the total angular momentum comes from the individual nucleons in a coherent superposition manner.

For all the present cesium isotopes, only one proton particle occupying at the bottom of the ${h}_{11/2}$ orbit contributes the angular momentum of roughly $5.5\ {\hbar}$ along the $s$ axis. It should be emphasized that the angular momentum contribution from proton part is approximately along the $s$ axis at low rotational frequency. Therefore, the angular momentum components along $l$ axis and $m$ axis are mainly from neutron part. At this point, ${^{124}{\textrm{Cs}}}$, ${^{125}{\textrm{Cs}}}$ and ${^{126}{\textrm{Cs}}}$ are used as the demonstration to study the angular momentum contributions of the valence neutrons in the ${h}_{11/2}$, $(gd)$ and $(sd)$ orbits along the $s$, $m$ and $l$ axis, which is shown in Fig.~\ref{fig4}. In contrast, five neutron holes locating at the top of the ${h}_{11/2}$ orbit contribute about $3.5\ {\hbar}$ to the angular momentum along the $l$ axis since four of them are antialigned. The only difference is the neutrons distributed over the $(gd)$ orbits and results in different configurations ${{\pi{{h}^{1}_{11/2}}}{\otimes}{\nu{{h}^{-1}_{11/2}}}{{(gd)}^n}{(n=12-14)}}$ for ${^{124-126}{\textrm{Cs}}}$.

Figure~\ref{fig4} shows that the angular momentum increment of the $(gd)$ orbits along the $s$ and $l$ axis will become smaller from ${^{124}{\textrm{Cs}}}$ to ${^{126}{\textrm{Cs}}}$ and the angular momentum increment along the $m$-axis will become larger. The similar behavior exists with the $(gd)$ or $(sd)$ orbits from ${^{120}{\textrm{Cs}}}$ to ${^{134}{\textrm{Cs}}}$. So, we speculate that the angular momentum increment of the $(gd)$ or $(sd)$ orbits along the $s$ and $l$ axis will become smaller with the increasing of the neutron number and the angular momentum increment along the $m$-axis will become larger. It is easier for the nucleus to form the chiral rotation, and the corresponding critical frequency ${\omega}_{\textrm{crit}}$ will decrease. Therefore, it is crystal clear that the critical frequency with the same configuration in Fig.~\ref{fig2} always tends to move to the left, which is similar to the earlier studies in Ref.~\cite{PhysRevC.105.044318}. It should be noted that the ${^{133}{\textrm{Cs}}}$, the $(sd)$ orbits neutron in this nucleus is transformed from the particle state character to the hole state behavior since a different change to others.

\begin{figure*}[ht]
\centering
\includegraphics[width=14cm]{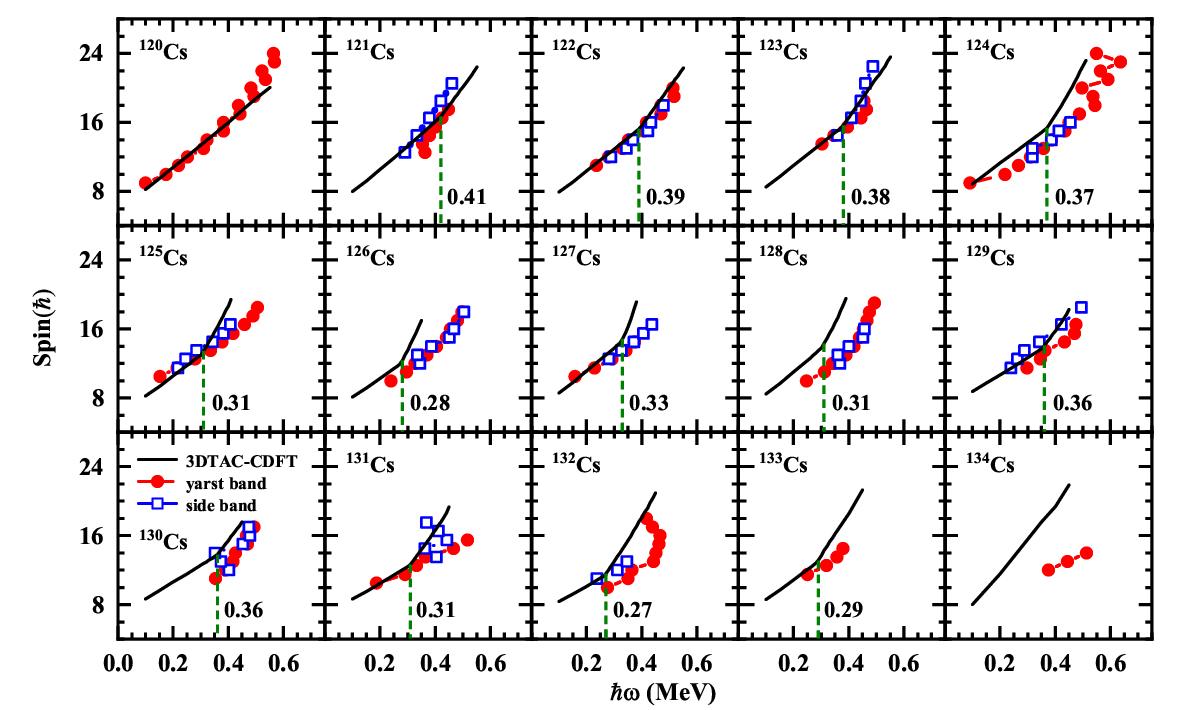}
\caption{\label{fig:exp1} (color online) The spin as functions of the rotational frequencies ${\hbar}{\omega}$ by $\textrm{3DTAC-CDFT}$ in comparisons with available experimental data~\cite{{PhysRevLett.86.971}, {PhysRevLett.97.172501}, {U_2005}, {GRODNER201146}, {Dong_2009}, {PhysRevC.96.051303(R)}, {PhysRevC.68.024318}, {Singh2006}, {PhysRevC.78.034313}, {PhysRevC.79.044317}, {PhysRevC.42.890}, {etde_20450471}, {Singh2005}, {PhysRevC.95.064320}, {PhysRevC.58.1849}, {PhysRevC.67.044319}, {PhysRevC.97.034306}, {PhysRevC.100.034328}} in $^{120-134}{\textrm{Cs}}$.}\label{fig5}
\end{figure*}

The critical frequency also affects the behaviors of experimental spin-rotational frequency $I(\hbar\omega)$ relationship. As shown in Fig.~\ref{fig5}, the calculated values of spin $I(\hbar\omega)$ agree well with the experimental data for $^{120-123}{\textrm{Cs}}$, and overestimate the data for $^{124-127,129-131,133}{\textrm{Cs}}$. In addition, the calculated values of spin $I(\hbar\omega)$ is much higher than the experimental data for $^{128,132,134}{\textrm{Cs}}$. For $^{124-134}{\textrm{Cs}}$, the effective neutron and proton pairing correlation and beyond mean-field effects which are not considered here may play the important roles, which has been confirmed in Ref.~\cite{{PhysRevC.101.054303}, {WANG2023137923}}. Furthermore, the back-bending phenomenon in $^{128,132,134}{\textrm{Cs}}$ which are not considered here may also affect the present calculated results much higher than the experimental data. Obviously, the calculated and experimental rotational frequencies ${\hbar}{\omega}$ for the yrast bands in $^{121-133}{\textrm{Cs}}$ behave similarly with respect to the angular momentum. The bands show nearly a straight line at low rotational frequency. A kink will appear when the spin reaches the critical frequency ${\omega}_\textrm{crit}$ in $^{121-133}{\textrm{Cs}}$. The occurrence of this kink is due to the transition from the planar to the chiral solutions and the angular momentum component will increase along the $m$ axis causing huge changes in the angular momentum. The corresponding kink of experimental $I({\hbar}{\omega})$ appears almost the same as the calculated ${\hbar}{\omega}_\textrm{crit}$ results. Further experimental efforts for $^{133}{\textrm{Cs}}$ are encouraged to verify the present conclusion on the ${\hbar}{\omega}_\textrm{crit}$. For $^{120,134}{\textrm{Cs}}$, previous experimental investigation does not support static chirality interpretation in the spin range observed~\cite{etde_20450471, PhysRevC.67.044319, PhysRevC.84.041301(R)}, which agrees with the theoretical result of no kink appearing.

\begin{figure*}[ht]
\centering
\includegraphics[width=14cm]{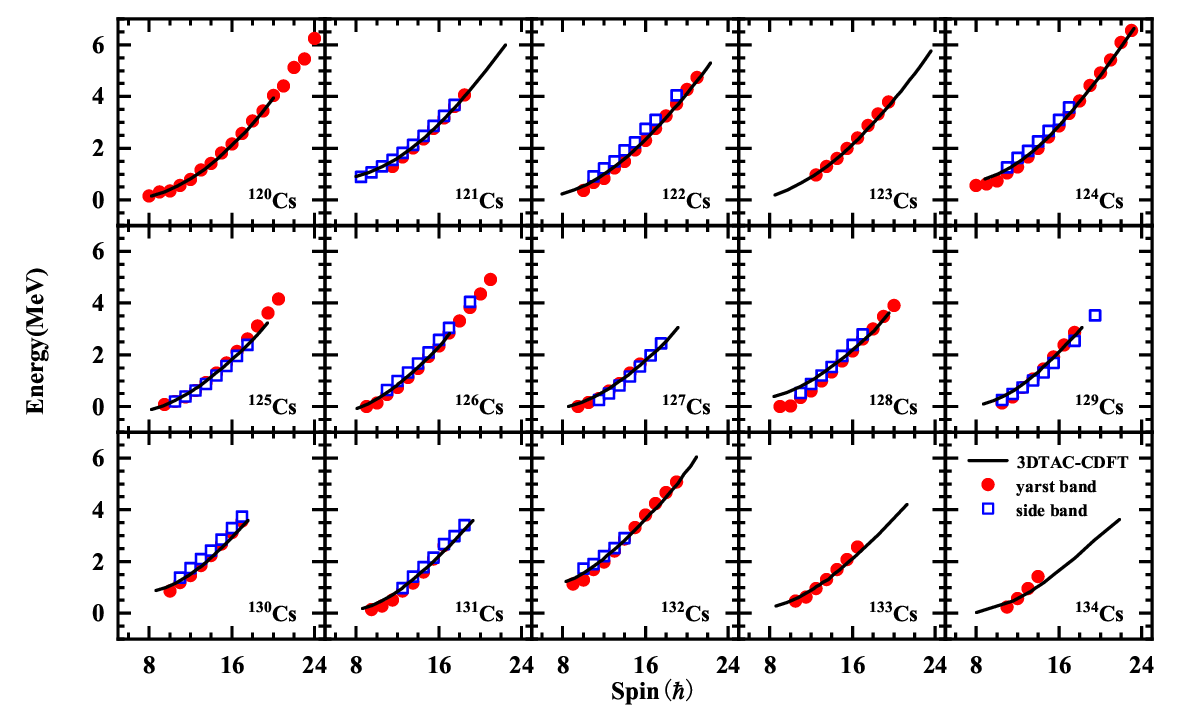}
\caption{\label{fig:energy} (color online) The calculated rotational excitation energies in $^{120-134}{\textrm{Cs}}$ by $\textrm{3DTAC-CDFT}$, as a function of the spin in comparison with the data for the chiral bands observed in Refs.~\cite{{PhysRevLett.86.971}, {PhysRevLett.97.172501}, {U_2005}, {GRODNER201146}, {Dong_2009}, {PhysRevC.96.051303(R)}, {PhysRevC.68.024318}, {Singh2006}, {PhysRevC.78.034313}, {PhysRevC.79.044317}, {PhysRevC.42.890}, {etde_20450471}, {Singh2005}, {PhysRevC.95.064320}, {PhysRevC.58.1849}, {PhysRevC.67.044319}, {PhysRevC.97.034306}, {PhysRevC.100.034328}}.}\label{fig6}
\end{figure*}

As aforementioned, the positive-parity rotational bands have been observed in $^{120-134}{\textrm{Cs}}$ in the previous works~\cite{{PhysRevLett.86.971}, {PhysRevLett.97.172501}, {U_2005}, {GRODNER201146}, {Dong_2009}, {PhysRevC.96.051303(R)}, {PhysRevC.68.024318}, {Singh2006}, {PhysRevC.78.034313}, {PhysRevC.79.044317}, {PhysRevC.42.890}, {PhysRevC.97.034306}, {PhysRevC.100.034328}}, and the comparisons between the 3DTAC-CDFT results of the excitation energies and the available experimental data are shown in Fig.~\ref{fig6}. In the present mean-field level, either the chiral vibrations or the tunneling between the left- and right-handed sectors are not taken into account. Therefore, only the band with lower excitation energies can be reproduced. However, it can be seen in Fig.~\ref{fig6} that the experimental excitation energies of the lower band can be reproduced well. To describe the partner band, one needs to go beyond the mean-field calculations by combining, for example, the methods of the random phase approximation~\cite{PhysRevC.83.054308} or the collective Hamiltonian with the CDFT~\cite{{PhysRevC.87.024314}, {PhysRevC.94.044301}}.

\begin{figure*}[ht]
\centering
\includegraphics[width=14cm]{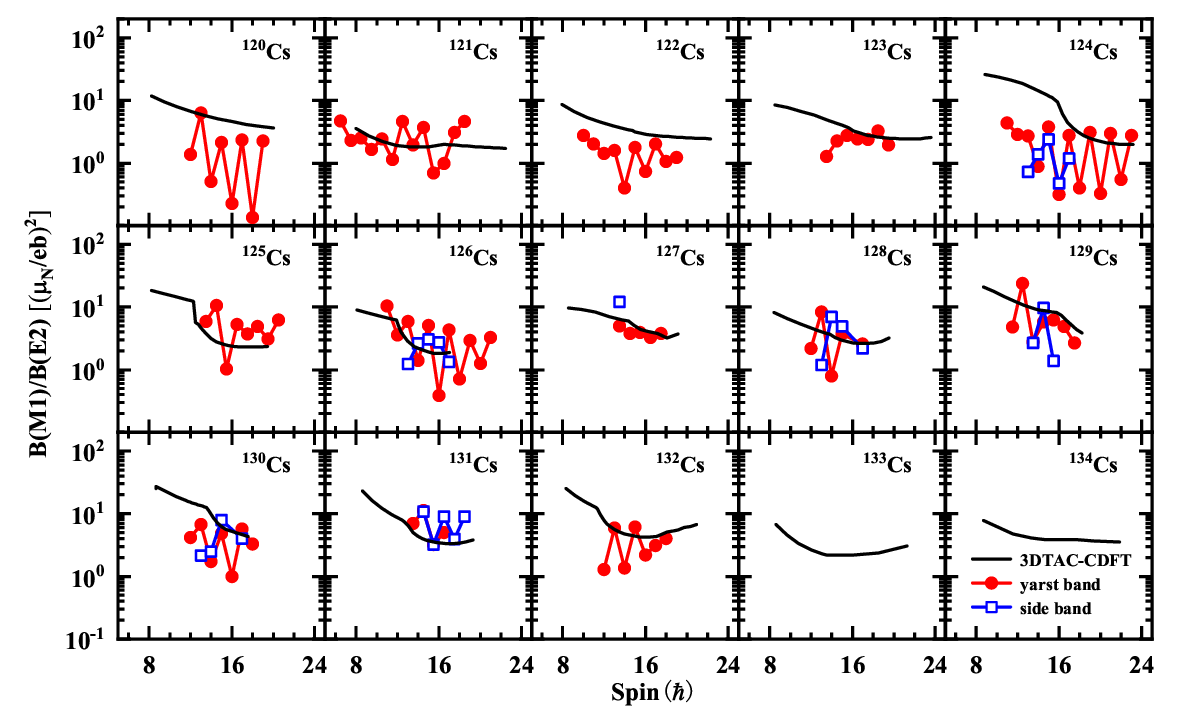}
\caption{\label{fig:B(M1)/B(E2)} (color online) The calculated $B(M1)/B(E2)$ ratios in $^{120-134}{\textrm{Cs}}$ by $\textrm{3DTAC-CDFT}$, as a function of the spin in comparison with the data for the chiral bands observed in Refs.~\cite{{PhysRevLett.86.971}, {PhysRevLett.97.172501}, {U_2005}, {GRODNER201146}, {Dong_2009}, {PhysRevC.96.051303(R)}, {PhysRevC.68.024318}, {Singh2006}, {PhysRevC.78.034313}, {PhysRevC.79.044317}, {PhysRevC.42.890}, {etde_20450471}, {Singh2005}, {PhysRevC.95.064320}, {PhysRevC.58.1849}, {PhysRevC.67.044319}, {PhysRevC.97.034306}, {PhysRevC.100.034328}}.}\label{fig7}
\end{figure*}

In Fig.~\ref{fig7}, the available experimental data of $B(M1)/B(E2)$ in $^{120-134}{\textrm{Cs}}$~\cite{{PhysRevLett.86.971}, {PhysRevLett.97.172501}, {U_2005}, {GRODNER201146}, {Dong_2009}, {PhysRevC.96.051303(R)}, {PhysRevC.68.024318}, {Singh2006}, {PhysRevC.78.034313}, {PhysRevC.79.044317}, {PhysRevC.42.890}, {PhysRevC.97.034306}, {PhysRevC.100.034328}} are displayed in comparison with the 3DTAC-CDFT results, which are derived in the semiclassical approximation from the magnetic and quadrupole moments. Note that the deformation parameters ${(\beta}, {\gamma)}$ change only slightly along the band, so the corresponding $B(E2)$ values are almost constant. However, the $B(M1)$ values decrease smoothly due to the continuous variation of rotational frequency and the closing of the neutron and proton angular momentum vectors, which mainly align along the short and long axes, respectively. This leads to the smooth-decreasing tendency for the $B(M1)/B(E2)$ ratios. Therefore, the ratios of $B(M1)/B(E2)$ by 3DTAC-CDFT are different from the experimental data, and only the smooth-decreasing tendency for the $B(M1)/B(E2)$ ratios can be given. In particular, the magnetic moments are derived from the relativistic expression of the effective current operator as in Ref.~\cite{RING1996193}. As shown in Fig.~\ref{fig7}, the 3DTAC-CDFT results for $^{121,123,125-134}{\textrm{Cs}}$ show good agreements with the data, while the calculation overestimates the data for $^{120,122,124}{\textrm{Cs}}$. The theoretical results show that $B(M1)/B(E2)$ for ${^{121-133}{\textrm{Cs}}}$ have a smooth falling behavior in the planar rotation, and the falling tendency slows down above the critical frequency ${\omega}_{\textrm{crit}}$. It once again proves that the nuclei will transition from the planar rotation to the chiral rotation above the critical frequency ${\omega}_{\textrm{crit}}$. The falling tendency shows steeper and steeper with increasing neutron number. Further efforts of including the neutron and proton pairing correlations will be helpful to justify the present results.

\begin{figure}[h]
\centering
\includegraphics[width=8cm]{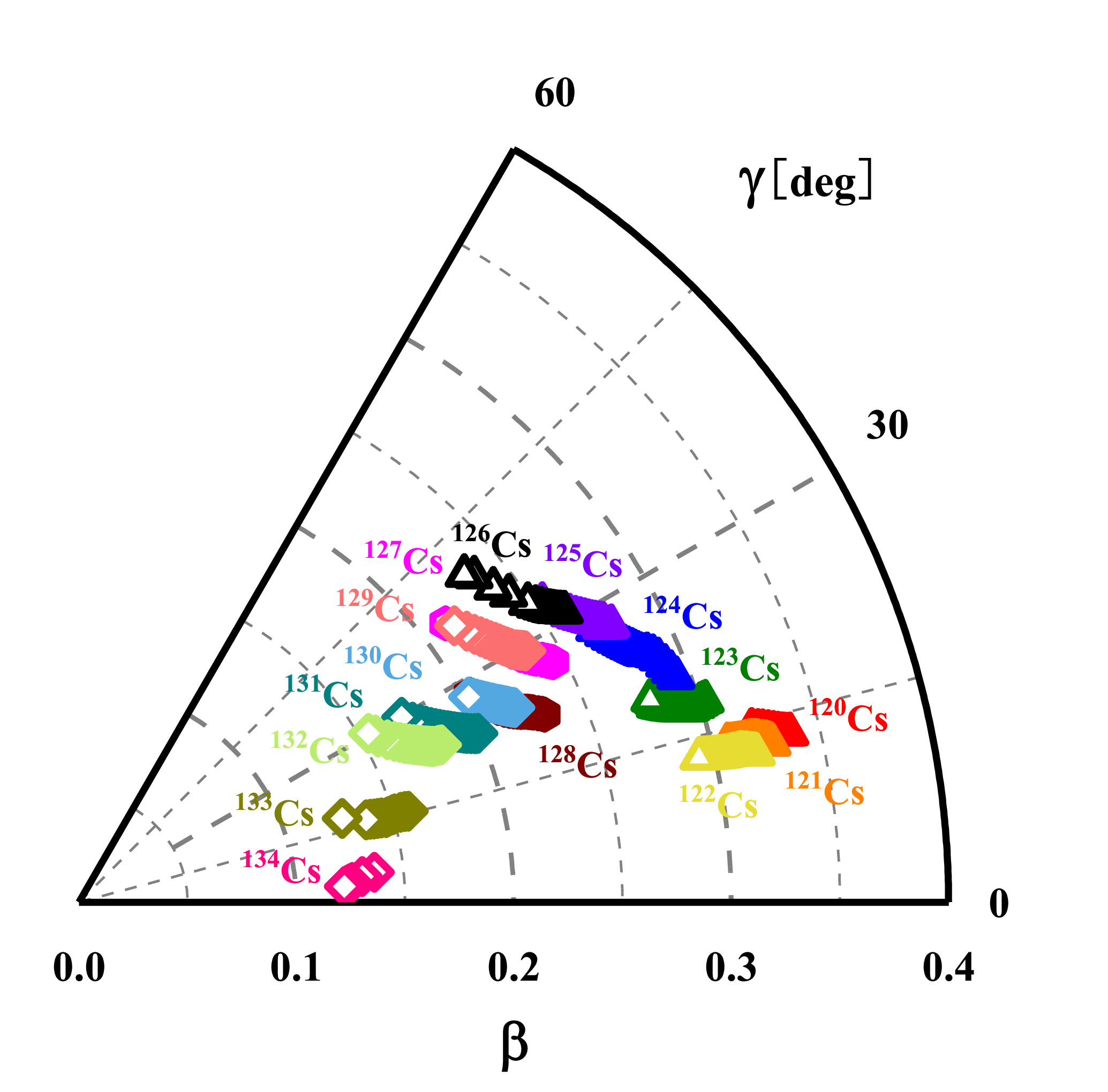}
\caption{\label{fig:deformation} (color online) The evolutions of the deformation parameters $\beta$ and $\gamma$ in the 3DTAC-CDFT calculations for ${^{120-134}{\textrm{Cs}}}$ with the configuration ${{\pi{{h}^{1}_{11/2}}}{\otimes}{\nu{{h}^{-1}_{11/2}}}{{(gd)}^n}}$ and ${{\pi{{h}^{1}_{11/2}}}{\otimes}{\nu{{h}^{-1}_{11/2}}}{{(sd)}^n}}$, respectively.}\label{fig8}
\end{figure}

The calculated quadrupole deformation parameters $\beta$ and triaxial deformation parameters $\gamma$ are given in Fig.~\ref{fig8}. It could be found that $\beta$ and $\gamma$ are stable and there is only one slight change with the increasing rotational frequency in ${^{120-134}{\textrm{Cs}}}$. With the increasing neutron number, the quadrupole deformation $\beta$ in cesium isotopes gradually decreases which is consistent with the results in Refs.~\cite{{MOLLER1995185}, {MOLLER20161}}. It could be understood that the number of neutrons in ${^{120}{\textrm{Cs}}}$ is the mediant of the magic number, so the quadrupole deformation $\beta$ will gradually decrease. For the triaxial deformation $\gamma$, it increases with the number of neutrons in ${^{120-126}{\textrm{Cs}}}$, and then decreases with the number of neutrons in ${^{127-134}{\textrm{Cs}}}$. It could be understood that the triaxial deformation $\gamma$ in ${^{120-126}{\textrm{Cs}}}$ gradually increases with the number of neutrons, the neutron holes contribution becomes stronger, and the deformation changes from prolate ellipsoid to oblate ellipsoid, which leads to the triaxial deformation $\gamma$ to grow up. Similarly, the neutron particle contribution becomes weakened with the increasing number of neutrons in ${^{127-134}{\textrm{Cs}}}$, and the deformation evolves from oblate ellipsoid to prolate ellipsoid, which leads the triaxial deformation $\gamma$ to decline. However, the suddenly increasing triaxial deformation $\gamma$ between ${^{128}{\textrm{Cs}}}$ and ${^{129}{\textrm{Cs}}}$ is due to the increasing neutron number in $h_{11/2}$ orbit.

It could also be found from Fig.~\ref{fig8} that the triaxial deformation in ${^{134}{\textrm{Cs}}}$ is approximately zero, and it further indicates there is no chirality in ${^{134}{\textrm{Cs}}}$, in agreement with the previous result~\cite{PhysRevC.67.044319}. The triaxial deformation parameter $\gamma$ in ${^{120}{\textrm{Cs}}}$ show relatively obvious value ($\sim 15^{\circ}$), although the planar rotation of zero azimuth angle $\varphi$ has been obtained. Therefore, the existence of chirality can not be completely obtained based on the high-$j$ particle-hole configurations and stable triaxial deformation~\cite{PhysRevC.97.034306}.

\section{Summary}

In summary, a fully self-consistent and microscopic investigation for the evolution of chirality in the cesium isotopes $^{120-134}{\textrm{Cs}}$ has been performed with the 3DTAC-CDFT. By investigating the evolution of the polar angle $\theta$ and azimuth angle $\varphi$ for the total angular momentum $J$ as driven by the increasing rotational frequency $\hbar\omega$, the transition from the planar rotation to the chiral rotation has been found in $^{121-133}{\textrm{Cs}}$. In comparison, only planar rotation has been obtained in $^{120,134}{\textrm{Cs}}$, and it is probably not the magnetic rotation.

Moreover, the critical frequency $\omega_{\textrm{crit}}$ in $^{121-133}{\textrm{Cs}}$ decreases with the increasing neutron numbers, and it also can be observed by the behaviors of $I$ and $B(M1)/B(E2)$ vs. $\hbar\omega$ relationship. Taking ${^{124}{\textrm{Cs}}}$, ${^{125}{\textrm{Cs}}}$ and ${^{126}{\textrm{Cs}}}$ as examples to study the angular momentum contributions of the valence neutrons in the ${h}_{11/2}$, $(gd)$ and $(sd)$ orbits along the $s$, $m$ and $l$ axis, the angular momentum increment of the $(gd)$ or $(sd)$ orbits along the $s$ and $l$ axis will become smaller with the increasing of the neutron number and the angular momentum increment along the $m$-axis will become larger. It is easier for the nucleus to form the chiral rotation, and the corresponding critical frequency ${\omega}_{\textrm{crit}}$ will decrease.

In addition, the calculated $I \sim \hbar\omega$ show good agreement with the available experimental data for $^{120-123}{\textrm{Cs}}$, and overestimate the data for $^{124-134}{\textrm{Cs}}$. The calculated energy spectra for yrast bands of $^{120-134}{\textrm{Cs}}$ are reasonable with the corresponding experimental data. The 3DTAC-CDFT results show a good agreement with the data of $B(M1)/B(E2)$ for $^{121,123,125-134}{\textrm{Cs}}$, while overestimates the data for $^{120,122,124}{\textrm{Cs}}$. The quadrupole deformation parameters $\beta$ and triaxial deformation parameters $\gamma$ in $^{120-134}{\textrm{Cs}}$ show obvious regular changes. The $\beta$ decreases in $^{120-134}{\textrm{Cs}}$ with the increase of neutron numbers. $\gamma$ increases in ${^{120-126}{\textrm{Cs}}}$ and decreases in ${^{127-134}{\textrm{Cs}}}$, which is related to the occupation of neutron numbers. Triaxial deformation in ${^{120}{\textrm{Cs}}}$ and ${^{134}{\textrm{Cs}}}$ indicates that the existence of chirality can not be completely obtained based on the high-$j$ particle-hole configurations and stable triaxial deformation.

Recently, the chiral bands of $^{119}{\textrm{Cs}}$~\cite{Zheng2022} based on configurations with only protons are observed for the first time in the ${A}\ {\sim}\ {130}$ mass region, i.e., a configuration with three protons, one in the strongly coupled $[404]9/2^+$ orbital which does not change orientation with increasing rotational frequency, and two in the $h_{11/2}$ orbital which reorients to the rotation axis, and it is different from the common chiral geometry picture from both protons and neutrons. It is an excellent significant and interesting work. The preliminlar results of 3DTAC-CDFT calculation don't support the exsitence of aplanar rotation with such configuration. At the same time, not only the TAC-CDFT, but also more other theoretical models such as particle-rotor model is helpful to justify the present results and understand the underlying rotational structure in $^{119}{\textrm{Cs}}$. Investigations in these directions are in progress. In addition, more efforts on the investigation of nuclear chirality in the ${A}\ {\sim}\ {130}$ mass region are still needed, including different theoretical methods, the evolution of the isotones, possible configurations and so on.

\begin{acknowledgements}

The authors would like to thank Prof. P. W. Zhao and Q. B. Chen for helpful discussions. This work was supported by the Natural Science Foundation of Jilin Province (No. 20220101017JC) and the National Natural Science Foundation of China (No 11675063) as well as the Key Laboratory of Nuclear Data foundation (JCKY2020201C157).

\end{acknowledgements}

\end{document}